\documentclass[conference]{IEEEtran}
\usepackage{pgfplots,pgfplotstable,booktabs}

\pgfplotsset{compat=1.5}

\usepackage{amsmath, amssymb}

\usepackage[figurename=Figure]{caption}

\usepackage{subfig}

\graphicspath{{images/}}

\usepackage{etoolbox}
\apptocmd{\sloppy}{\hbadness 10000\relax}{}{}

\usepackage{cite}

\begin{document}

\title{Appearance of multiple stable load flow solutions under power flow reversal conditions}
%Integrated PV Distribution Networks: Stability of New Solutions}

\author{\IEEEauthorblockN{Hung D. Nguyen}
\IEEEauthorblockA{School of Mechanical Engineering\\
Massachusetts Institute of Technology\\
Cambridge, MA 02139\\
Email: hunghtd@mit.edu}
\and
\IEEEauthorblockN{Konstantin S. Turitsyn}
\IEEEauthorblockA{School of Mechanical Engineering\\
Massachusetts Institute of Technology\\
Cambridge, MA 02139\\
Email: turitsyn@mit.edu}}

% make the title area
\maketitle

\begin{abstract}
\boldmath
In complex power systems, nonlinear load flow equations have multiple solutions. Under typical load conditions only one solution is stable and corresponds to a normal operating point, whereas the second solution is not stable and is never realized in practice. However, in future distribution grids with high penetration of distributed generators more stable solutions may appear because of active or reactive power reversal. The systems can operate at different states, and additional control measures may be required to ensure that it remains at the appropriate point. This paper focuses on the analysis of several cases where multiple solution phenomena is observed. A non-iterative approach for solving load flow equations based on the Gr\"{o}bner basis is introduced to overcome the convergence and computational efficiency associated with standard iterative approaches. All the solutions of load flow problems with their existence boundaries are analyzed for a simple 3-bus model. Furthermore, the stability of the solutions is analyzed using a derived aggregated load dynamics model, and suggestions for preventive control are proposed and discussed. The failure of na\"{i}ve voltage stability criteria is demonstrated and new voltage stability criteria is proposed. Some of the new solutions of load flow equations are proved to be stable and/or acceptable to the EN 50610 voltage fluctuation standard. 
\end{abstract}


\IEEEpeerreviewmaketitle
%Introduction
\section{Introduction}

Modern power grids are composed of millions of highly nonlinear devices interconnected into one system. The operating point of the power system is a solution of a highly nonlinear system of equations known as the load flow problem \cite{Kirtley}. Even for fixed levels of power consumption and generation multiple solutions of load flow equations can coexist. However, under typical conditions, not more than one solution is stable and compliant with the voltage standards. This solution is referred to as the normal operating point \cite{Klos1991268}, \cite{trias}. The normal solution is characterized by relative high voltage level and correspondingly low current. In most of the systems, the normal solution is also the only stable solution. The advent of distributed generators (DGs), either renewable or gas-fired, will cause the distribution grids to operate in currently uncommon conditions. In particular, the flow of active or reactive power may become reversed.  The conditions for stability and the existence of a solution in these grids may be very different, new stable solutions may appear. The existing voltage protection and control systems may no longer ensure stable and secure operation of the distribution grid. Novel approaches may be required if there are several solutions that are both stable and compliant with voltage level standards.

Analysis of all the solution branches of load flow equations is a computationally challenging problem. The most commonly used iterative methods suffer from divergence problem in the vicinity of the bifurcation. The common causes for the divergence are: poor initial guesses, small distance from the solution existence boundary, and nonexistence of the solutions \cite{JEET}. Other numerical approaches, like continuation methods may overcome the divergence problem but can not easily identify all the solution branches \cite{Kundur}. An alternative approach based on the algebraic geometrical technique of the Gr\"{o}bner basis was proposed in \cite{montes1,ning,kavasseri} but did not ever become mature enough for practical applications. In this work, we employ the algebraic geometry approaches to study the solution space of load flow problem. Unlike the iterative methods, the algebraic approach allows us to characterize the whole manifold of solutions at once. This leads to dramatic reductions in computational efforts.

Existence of multiple solutions may pose additional risks on the system and jeopardize the voltage security of the system. In particular, existing protection against voltage collapse phenomenon may become inadequate for new conditions. Voltage collapse is the process in which the sequence of events accompanying voltage instability leads to a blackout or abnormally low voltages in a significant part of the power system \cite{Definition}. Voltage collapse has been deemed responsible for several major blackouts \cite{pes2012}. Unfortunately, no simple criterion is known to the power systems engineers that would allow to test the voltage stability given a static load flow solution \cite{Yokoyama}. Dynamic information about load behavior should be incorporated in the model and there have been a large number of studies that proposed various criteria for voltage stability based on a number of dynamic models \cite{Cutsem,DrHill,Iwamoto}.

The structure of our paper is organized as follows. In part \ref{sec:groebner} of the paper we introduce the Gr\"{o}bner basis technique for characterization of the solution manifold. Then, in subsections \ref{sec:solution} and \ref{sec:number}, this technique is applied to three-bus system and the phase diagram of the number of solution branches is derived and analyzed. We derive a novel voltage stability criterion in section \ref{sec:stability}. The results are applied in numerical simulations in section \ref{sec:simulation}.

\section{The Gr\"{o}bner basis applications}\label{sec:groebner}

\subsection{Characterizing solution branches}

Several techniques have been introduced in the literature for identification of all power flow solutions, the optimal multiplier based method \cite{iwamotooptimal}, and more recently a holomorphic embedding method \cite{trias}. As aforementioned, an alternative approach is based on the Gr\"{o}bner basis technique applicable to any systems of polynomial equations. The  introduction to the Gr\"{o}bner basis approach can be found in \cite{Cox}, \cite{Buchberger}, while here we introduce the well-known Buchberger's algorithm for solving the set of polynomial equations.
%\cite{}[Ilic, Okumura, Antonio Trias, US patent]

%All power flow solutions could be found by either choosing multiple initial conditions \cite{}, for instance, 4 appropriate initial conditions could be used to find 4 respective solutions, or using holomorphic method \cite{}[Ilic, Okumura, Antonio Trias, US patent]. 

\begin{figure}[ht]
    \centering
    \includegraphics[width=8cm]{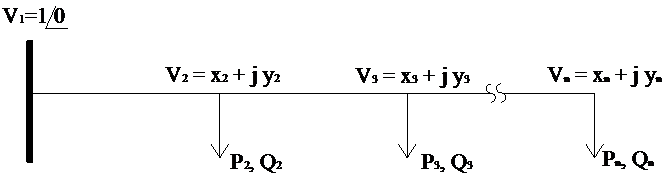}
    \captionof{figure}{A radial network}
	%\caption{A radial network}
    \label{nbuslink}
\end{figure}

We use the rectangular form of power flow equations. For example, consider a radial link with $n$ buses is shown in Fig. \ref{nbuslink}. In this grid, the bus 1 is slack bus with voltage $ V_1=1\angle{0} $. 

At bus $i$, $ 2\leq i \leq n $, let i) $ P_i $ and $ Q_i $ be active and reactive power respectively; ii) $ V_i=V_{iRe}+jV_{iIm} $ be the rectangular form of bus voltage. So power flow equations for PQ buses can be expressed as follows \cite{kavasseri}:
\begin{eqnarray} 
\begin{split} \label{eq:pfcom}
P_i & = \sum\limits_{k=1}^n G_{ik}(V_{iRe}V_{kRe}+V_{iIm}V_{kIm})\\
& +\sum\limits_{k=1}^nB_{ik}(V_{kRe}V_{iIm}-V_{iRe}V_{kIm}); \\
Q_i &= \sum\limits_{k=1}^n G_{ik}(V_{kRe}V_{iIm}-V_{iRe}V_{kIm})\\
& -\sum\limits_{k=1}^nB_{ik}(V_{iRe}V_{kRe}+V_{iIm}V_{kIm})  
\end{split}
\end{eqnarray}
where $Y_{ik}=G_{ik}+jB_{ik} $ is an entry of the bus admittance matrix, $Y$.

Application of Buchberger's algorithm in Mathematica allows one to transform the system of equations in a new ``triangular'' form:

\begin{equation} \label{eq:buchergersol}
 \sum\limits_{k=0}^{2^{n-1}}a_kV^k_{n\, Im} = 0     \nonumber
\end{equation}

\begin{equation}
\sum\limits_{k=0}^{2^{n-1}}b_kV^k_{n\,Im}+A_nV_{nRe} = 0  \nonumber
\end{equation}

\begin{equation}
\sum\limits_{k=0}^{2^{n-1}}c_kV^k_{n\,Im}+B_{n-1}V_{n-1\,Im} = 0  
\end{equation}

\begin{equation}
\sum\limits_{k=0}^{2^{n-1}}d_kV^k_{n\,Im}+A_{n-1}V_{n-1\,Re} = 0 \nonumber
\end{equation}

\[
\dots
\]
The main advantage of this form is that the equations can be solved subsequently. The first equation of (\ref{eq:buchergersol}) is first solved for $ V_{nIm} $. Substituting $ V_{nIm} $ into the second equation of (\ref{eq:buchergersol}) allows to solve it for $ V_{nRe} $. The iteration procedure continues until $ V_{2Re} $ is found. Finally, the voltage magnitude of any bus can be determined as:
\begin{equation}
|V_i|=\sqrt{V^2_{iRe}+V^2_{iIm}} \label{eq:volmag}
\end{equation}
The maximum number of solutions of each equation of (\ref{eq:buchergersol}) is $ 2^{n-1} $ and that determines the maximum number of solutions of the radial link. For the 3-bus radial link, obviously, at maximum 4 solutions can be founded.

%For $n=3$, we consider a 3-bus system as Fig. \ref{3buslink}. 
%\begin{figure}[ht]
%   \centering
%   \includegraphics[width=0.35 \textwidth]{3bus}
%	\caption{A 3-bus radial network}
%   \label{3buslink}
%\end{figure}

\subsection{Finding Solution Boundaries}\label{sec:solution}

\subsubsection{Numerical method} In \cite{hiskensbound}, Hiskens et al. have proposed a technique to studty the solution space space boundaries using the following set of equations:
\begin{equation} %\label{eq:hk1}
f(x,\lambda) = 0 \nonumber
\end{equation}
\begin{equation} \label{eq:hk}
f_x(x,\lambda)v = 0
\end{equation}
\begin{equation} %\label{eq:hk3}
v^tv = 1\nonumber
\end{equation}
% where $x$ is state vector of the system, $x \in \Re^{2(n-1)} $; $ \lambda $ is a parameter which is free to vary; $ f(x,\lambda)= 0 $ is the set of power flow equations; $ f_x $ is the Jacobian matrix $ [\partial{f}/\partial{x}] $; $v$ is the right eigenvector corresponding to a zero eigenvalue of the Jacobian matrix, $ f_x $, $v \in \Re^k $. 

Here $v$ is the right eigenvector of the Jacobian matrix corresponding to its zero eigenvalue. If the total number of parameters is equal to $k$, so that $\lambda \in \mathbb{R}^k$, and both the vectors $x$ and $v$ have $2(n-1)$ components, so that $x,v\in \mathbb{R}^{2n-2}$, the total number of variables in this equation is $k+4n-4$. The total number of equations on the other hand is $4n-3$, so the system (\ref{eq:hk}) describes the $k-1$ dimensional manifold in the parameter space $\lambda$. 

\subsubsection{Symbolic computation} 
\begin{figure}[ht]
    \centering
    \includegraphics[width=0.45\textwidth]{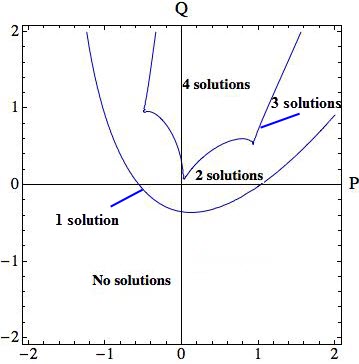}
	\caption{A three-bus network's solution boundaries}
    \label{solbound}
\end{figure}

In the Gr\"{o}bner basis approaches, in order to determine the solution space boundaries of the load flow equations using, we also use (\ref{eq:hk}) with $x=(V_{Re},V_{Im})^t$ where $V_{Re}=(V_{2Re},V_{3Re},...,V_{nRe})$; $V_{Im}=(V_{2Im},V_{3Im},...,V_{nIm})$; and $\lambda=(P_i,...,Q_j)^t$, $ \lambda \in \Re^k $. For (\ref{eq:hk}), Buchberger's algorithm gives us algebraic equations of $P$, and $Q$, for instance, $g(P,\,Q)=0$, that also provides the equation for the boundary manifold of the solution space. For the 3-bus system, the contour of solution space is plotted in Fig. \ref{solbound} where $\lambda=(P_2,Q_2)^t$, $P_2=P_3$, $Q_2=Q_3$. Since $k=2$, i.e. two free parameters, the set of equations of (\ref{eq:hk}) defines curves. Negative values of active and reactive powers indicate that the system is consuming active power with a lagging power factor, or \emph{vice versa}, positive values mean exporting active power with a leading power factor.

% There are $4n-3$ equations in the set of equations of (\ref{eq:hk}). The number of unknowns are $4(n-1)$ unknowns $V_{Re}, V_{Im}$, and $k$ unknowns $\lambda_i$ which are free to vary. Hence, the contour of the solution space is a $k-1$ manifold.

Fig. \ref{solbound} illustrates that the system may have multiple solutions even in normal operating conditions. Most interestingly, if the system exports a small amount of active and reactive power, 4 solutions can coexist. 
%In addition, the solution boundaries contradicts Okumura`s statement in \cite{okumura} that the power flow equation tends to have multiple solutions as the loads become heavier.
%contingency conditions

As was argued in \cite{hsiao-dongchiang} and confirmed by our analysis in most of the situations when multiple solutions coexist, all of the solutions but one correspond to low voltage profiles. Hence, most of the solutions branches can never be realized in normal conditions. However, in post-disturbance or heavily loaded conditions the system may get attracted to one of the unconventional branches due to nonlinear transient dynamics.  
%In order to emphasize the relevance of multiple high voltage solutions, angular instability is ignored. The fact is that unstable high voltage solutions are typically related to a shift of angles around $180^o$ \cite{castro}. In other words, the systems are subjected to an angle instability.
The phase diagrams similar to the ones depicted in Fig. \ref{solbound} can be used by system operators to identify the safe operation regions and help in diagnosing the system problems in emergency and post-emergency conditions. They can also be helpful in desigining various control systems, such as photovoltaic panel inverter control \cite{turitsyn2011options}, \cite{farivar2011inverter} or preventive control for restoring the systems' solvability \cite{overbye1994power}. 
%For a given power system, the actual solution boundaries are informative in the senses of telling where the system should work and how many solutions the system could have. They could work as a map which enables system operators to have a better sense of the system. One of the most important applications could be restoring the system's solvability, i.e. preventive control.
\subsection{Number of solution analysis}\label{sec:number}
\begin{figure}%
    \centering
    \subfloat[{Q2-V2 curve}]{{\includegraphics[width=4.1cm]{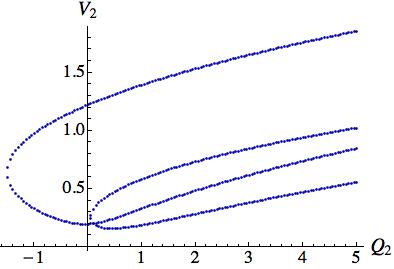} }}%
    \quad
    \subfloat[{P2-V2 curve}]{{\includegraphics[width=3.9cm]{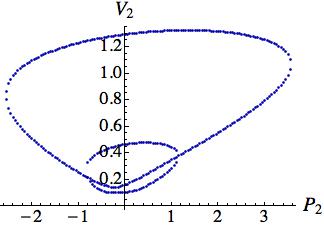} }}%
    \caption{Q2-V2 and P2-V2 curves}%
    \label{PQV}%
\end{figure}

The total number of solutions in radial network with PQ buses is always less than $2^{n-1}$, however it can be larger in transmission grids with PV buses and non-radial structure \cite{Klos1991268}. To illustrate the structure of new solutions we may look at the dependence between $Q_2, P_2$ and $V_2$ as shown on figures \ref{PQV}(a) and \ref{PQV}(b). With $P_2=0.5$, $P_3=0.5$, $Q_3=1$ (p.u.), the $Q_2-V_2$ curve is shown in Fig. \ref{PQV}(a). Each value of $Q_2$ greater than $0.12$ p.u. will result in 4 values of $V_2$. Hence, the system has 4 solutions. Otherwise, the system has either 2 solutions or no solutions. With $Q_2=0.5$, $P_3=0.5$, $Q_3=1$ (p.u.), the $P_2-V_2$ curve is shown in Fig. \ref{PQV}(b). Each value of $P_2$ between -1.13 p.u. and 1.21 p.u. will also result in 4 values of $V_2$.

%Our results are consistent with the more general result reported in \cite{Klos1991268} that shows $d\leq2^{n-1}+m$ where $d$ is the distinct solutions quantity, $m$ is a number which depends on the system configuration \cite{Klos1991268} and for radial power networks with no $P$, $|V|$ buses, $m=0$. So the maximum number of different solutions the system may have is $2^{n-1}$.

\subsection{Preventive Control}

In 1994-1995, Overbye presented a method for determining system controls in order to restore the power flow based on a damped Newton-Raphson power flow algorithm and a sensitivity analysis \cite{overbye1994power}, \cite{overbye1995computation}. This method is able to determine the solvable boundary and the best direction to shed the loads to restore solvability. This method works best when the solvable boundary is nearly flat however can lead to high errors in highly curved boundary situations. Other methods that minimize load shedding have been proposed \cite{JEET}, \cite{sangsoo}. However, all previous methods to our knowledge were based on numerical approaches. The advantage of the proposed algebraic approach is that explicit characterization of the solvable boundary could simplify the task of finding optimal preventive control. The control action in this case would move the system from the no solution region to the solvable region. Once the system stays in the solvable region, other voltage regulation methods could be applied in order to correct the system voltages such as steady state voltage monitoring and control \cite{JEET}, \cite{monvolilic}.

%\section{GENERALIZED VOLTAGE STABILITY CRITERIA}
\section{Generalized voltage stability criteria}

\subsection{Load dynamics}
The driving force for voltage instability is usually the load dynamics. Traditional loads restore the power levels in response to a disturbance via motor slip adjustment, distribution voltage regulators, tap-changing transformers and thermostats. Restored loads increase the stress on the high voltage network by increasing the reactive power consumption and causing further voltage reduction \cite{Definition}. Based on the load dynamic characteristics in \cite{Cutsem} and \cite{smartgrid}, we propose the following universal model for the load dynamics:
\begin{equation} \label{eq:loaddyn}
\dot{y}=-f(|V|^2)(p-P^0)
\end{equation}
where: $y$ is the load admittance magnitude, $y=g+jb$; $|V|$ is the voltage magnitude at load bus that varies as the load is attempting to regulate its demand to achieve a desired demand, $f(|V|^2)$ is some positive function, i.e. $f(|V|^2)>0$; $p$ is the load consumption which is regulated to achieve a desired demand, $P^0$.

\subsection{Small signal stability criteria}\label{sec:stability}
From (\ref{eq:pfcom}) and with fixed reactive powers, and active power demands, $p_i$, where $p_i \neq P^0$, Buchberger's algorithm could give us the solution in terms of active power consumptions:
\begin{equation} \label{eq:v-p}
|V_k|^2=|V_k|^2(p_1,p_2,...,p_n)
\end{equation}
Let $U_k=|V_k|^2$; $y_i=|y_i|$; $i=1,n$ for convenience.
The equation for voltage dynamics can be expressed as follows:
\begin{equation} \label{eq:dotv-y}
\dot{U_k}=\sum\limits_{j=2}^n\frac{\partial{U_k}}{\partial{y_j}}\dot{y_j}
\end{equation}
Without loss of generality, assume that $q_k=0$, thus $p_k=y_kU_k$; we have:
\begin{equation} \label{eq:Vk-pl}
\frac{\partial{U_k}}{\partial{y_j}}=\sum\limits_{l=2}^n\frac{\partial{U_k}}{\partial{p_l}}(\delta_{jl}U_l+y_l\frac{\partial{U_l}}{\partial{y_l}})
\end{equation}
where $\delta_{jl}=1$ if $j=l$, otherwise $\delta_{jl}=0$. (\ref{eq:Vk-pl}) becomes:
\begin{equation}
\sum\limits_{l=2}^n(\delta_{jl}-y_l\frac{\partial{U_l}}{\partial{p_l}})\frac{\partial{U_k}}{\partial{y_l}}=\frac{\partial{U_k}}{\partial{p_j}}U_j
\end{equation}
For convenience, define $\mathbf{U_y}=[\frac{\partial{U_k}}{\partial{y_j}}]$; $\mathbf{U_p}=[\frac{\partial{U_k}}{\partial{p_j}}]$. If $[1-\mathbf{U_p}diag(\frac{p}{U})]$ is nonsingular, we can compute $\mathbf{U_y}$ as:
\begin{equation} \label{eq:Vymatrix}
\mathbf{U_y}=[1-\mathbf{U_p}diag(\frac{p}{U})]^{-1}\mathbf{U_p}diag(U)
\end{equation}
The voltage sensitivity matrix, $\mathbf{U_y}$, has a very natural interpretation. Its entries express the sensitivity of bus voltages with respect to load admittances. In other words, each entry shows how the voltage at one bus would respond to the change of the load admittance at another bus.
The equation (\ref{eq:dotv-y}) can be written as:
\begin{equation} \label{eq:dotv-yexpan}
\dot{U_k}=-\sum\limits_{j=2}^nU_y(k,j)f(U_j)(p_j-P^0_j)
\end{equation}
where $\mathbf{U_y}(k,j)$ is the $(k,j)$ element of $\mathbf{U_y}$.
The equilibrium point is the solution of the following set of equations:
\begin{eqnarray} \label{eq:equi}
& \dot{U_k}=0\\
& Re(\sum\limits_{l=1}^nV^*_ky_{kl}V_l)-p_k=0
\end{eqnarray}
%In order to linearize conveniently around the equilibrium point, let $\dot{|y|}=-f(U)(U-U_0)$.
For the special case of slow load recovery, i.e. $p_k-P^0_k=0$, we can obtain:
\begin{equation} \label{eq:linearize}
\dot{\delta{U_k}}=-\sum\limits_{j=2}^nU_y(k,j)\,f(U)\,(\partial{p_j}/\partial{U_j})\,\delta{U_j}
\end{equation}
Therefore, the small signal stability at the equilibrium point is determined by the matrix:
\begin{equation} \label{eq:Jaco}
\mathbf{J}=-\mathbf{U_y}\,diag(f(U))\,\mathbf{p_U}
\end{equation}
where $\mathbf{p_U}=[\frac{\partial{p}}{\partial{U}}]=\mathbf{U_p}^{-1}$.
Hence, all eigenvalues of matrix $\mathbf{J}$ must be in the open left-half plane in order to ensure the system's small signal stability. This result is another representation of the small-signal stability criteria such as derived in \cite{Jiansun,DrHill,Iwamoto}.

\section{Simulations}\label{sec:simulation}

\subsection{Multiple high voltage solutions}

Consider a 3-bus network which is a modified one based on the IEEE Standard 4-Node Test Feeder \cite{IEEE4node} with the following conductance matrix and susceptance matrix:

$
\centering{
G = \begin{bmatrix} 2.04&-1.02\\ -1.02&1.02 \end{bmatrix} \quad
B = \begin{bmatrix} -7.17&3.59\\ 3.59&-3.59 \end{bmatrix}}
$

\begin{figure}[t]
    \centering
    \includegraphics[width=6cm]{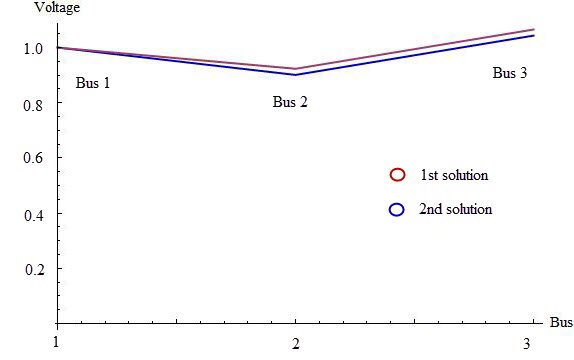}
	\caption{Multiple high voltage solutions}
    \label{multiphighsol}
\end{figure}

With $P^0_2=-1.105$, $Q^0_2=1$, $P^0_3=-1$, $Q^0_3=1.105$ (p.u.), multiple high voltage solutions are found as shown in Fig. \ref{multiphighsol} and Table \ref{soltable}. As one can see, multiple solutions can be observed in situations where the distribution feeder exports reactive power to the transmission grid. Under these conditions, the two solutions could be almost identical, i.e. having nearly the same magnitudes and angles. More importantly, both solutions are stable and acceptable according to the EU standard, EN 50160, for distribution networks, i.e. the fluctuation in voltage is less than $\pm10 \%$ of the nominal voltage \cite{EN50160}. Moreover, the set of load demands should be re-selected in order to obtain a better solution in terms of smaller voltage fluctuation range such as $\pm5 \%$ as the US standards.

\begin{figure}%
    \centering
    \subfloat[{P2-V2 curve}]{{\includegraphics[width=4cm]{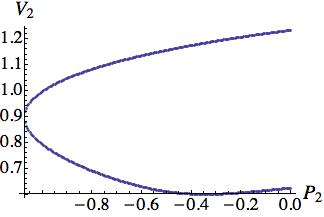} }}%
    \quad
    \subfloat[{Q2-V2 curve}]{{\includegraphics[width=4cm]{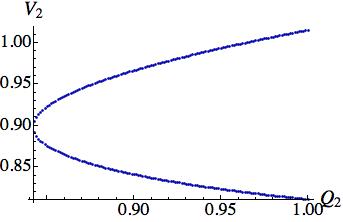} }}%
    \caption{The active and reactive power nose curves at bus 2}%
    \label{nose}%
\end{figure}

\begin{table}[ht]
    \caption{HIGH VOLTAGE SOLUTIONS}
    \centering
    \label{soltable}
    \begin{tabular}{|c |c |c|}
    \hline
    \textbf{Solution/Bus} & \textbf{Bus 2} & \textbf{Bus 3} \\
               \hline
    $1$st solution & $0.901\angle{-0.886}$ & $1.043\angle{-0.857}$ \\
    $2$nd solution & $0.923\angle{-1.255}$ & $1.065\angle{-1.209}$ \\
    \hline
    \end{tabular}
\end{table}
The similarity between the two solution is explained by the proximity of the system parameters to the solution space boundary. Like in the classical nose curve the two solution get close to one another as the load level approaches the critical value as shown in  Fig. \ref{nose}(a). The similar nose curve for reactive power at bus 2 is also depicted in Fig. \ref{nose}(b). This phenomenon illustrates the drawbacks of the reactive power compensation with DGs such as photovoltaic panels \cite{turitsyn2010local}. Although, the reactive power compensation can stabilize the voltage levels in the system, it can also cause the voltage collapse phenomenon to occur at the high voltage profile situation compliant with the existing standards. 

%Since PV are able to inject reactive power into the networks, the nose curves can be raised along the voltage axis. This implies an unexpected effect of PV, i.e. voltage collapse can happen even the system works with a standard voltage profile.
%normal operating conditions.

\subsection{Stability of new solutions}
We assume the following simple model for the dynamics of the two loads: $\dot{y_1}=-(p_1-P^0_1)$ and $\dot{y_2}=-2\,(p_2-P^0_2)$. For the above 3-bus system with $P^0_2=1.1$, $Q^0_2=1.8$, $P^0_3=1.5$, $Q^0_3=0.9$ ($p.u.$), and for the first solution $|V_2|=0.504$, $|V_3|=0.693$ ($p.u.$), we have:

$
\centering{
\mathbf{U_p}=\begin{bmatrix} -0.047&0.353\\ -0.995&1.093 \end{bmatrix} \quad
\mathbf{U_y}=\begin{bmatrix} -0.135&0.092\\ -0.136&-0.054 \end{bmatrix}}
$

\begin{equation}
\centering
\mathbf{J}=\begin{bmatrix} -0.115&-0.130\\ 0.854&-1.177 \end{bmatrix}\nonumber
\end{equation}
The eigenvalues of the Jacobian matrix are: $\lambda_{1,\,2}=-0.146\pm j\,0.332$. Therefore, the two eigenvalues of the Jacobian matrix are in the open left-half plane. So the first solution is stable. Similarly, for the fourth solution $|V_2|=1.491$, $|V_3|=1.683$ ($p.u.$), the eigenvalues of the Jacobian matrix are $\lambda_1=-5.460$, $\lambda_2=-2.201$. Hence, the second solution is also stable. This result agrees with our previous finding in \cite{turitsyn}. However, the coexistence of two stable solutions can not be observed in traditional networks where both active and reactive powers are consumed. However, whenever such a situation occurs a new interesting question should be posed: which of the two solutions is best? We will leave it for future research.

\section{Conclusion}
In this work we have shown that distribution grids with active or reactive power flow reversal can have multiple stable solutions of the load flow problems. The system may get attracted to either of those solutions as a result of transient dynamics and additional controls should be introduced to operate the system in those conditions. The properties of these solutions also need to be analyzed in more details. In our work we proposed a technique of characterizing the solution space boundary via the Gr\"{o}bner basis approach. Other applications of the Gr\"{o}bner basis approach have been also briefly reviewed. The stability of the new solutions has been checked with the derived generalized voltage stability criterion based on the conventional load dynamics model. We show that the distribution grids with high penetration of DGs with reactive power compensation capabilities may not be protected against voltage collapse with traditional voltage security measures. 

In the future we plan to extend the Gr\"{o}bner basis approach to large systems with the help of reduction and approximation techniques. Also, we are planning to use the bus voltage dynamics in (\ref{eq:dotv-y}) to derive the reduced models of distribution grids, and study the effect of distributed reactive power compensation on the overall stability of the system. Finally, we are planning to continue our work on the characterization and their potential uses for power systems. 

% use section* for acknowleement
\section*{Acknowlement}

The authors would like to thank the NSF foundation, MIT/SkTech initiative, and Vietnam Education Foundation for their support. 

\bibliographystyle{IEEEtran}
\bibliography{main.bll}
\end{document}